\documentclass{PoS}

\usepackage{bm,slashed}
\usepackage[caption=false]{subfig}

\graphicspath{{./Figures/}}

\newcommand{\diff}{\mathrm{d}}
\newcommand{\p}{\partial}

\newcommand{\up}{\uparrow}
\newcommand{\down}{\downarrow}
\newcommand{\be}{\begin{equation}}
\newcommand{\ee}{\end{equation}}
\newcommand{\bea}{\begin{eqnarray}}
\newcommand{\eea}{\end{eqnarray}}
\newcommand{\im}{\mathrm{i}}

\newcommand{\eqref}[1]{(\ref{#1})}

\title{Lefschetz-thimble approach to the Silver Blaze problem of one-site fermion model}

\ShortTitle{Lefschetz-thimble approach to the Silver Blaze problem of one-site fermion model}

\author{\speaker{Yuya Tanizaki}\\
     RIKEN BNL Research Center, Brookhaven National Laboratory, Upton, NY 11973 USA\\
        E-mail: \email{yuya.tanizaki@riken.jp}}

\author{Yoshimasa Hidaka\\
	Theoretical Research Division, Nishina Center, RIKEN, Wako 351-0198, Japan\\
	E-mail: \email{hidaka@riken.jp}}

\author{Tomoya Hayata\\
	Department of Physics, Chuo University, Tokyo 112-8551, Japan\\
	E-mail: \email{hayata@phys.chuo-u.ac.jp}}

\abstract{The sign problem of finite-density QCD at the zero temperature becomes very severe if the quark chemical potential exceeds half of the pion mass. 
In order to understand its property, we consider the sign problem of the one-site fermion model appearing in its path-integral expression by using the Lefschetz-thimble method. 
We show that the original integration cycle becomes decomposed into multiple Lefschetz thimbles at a certain value of the fermion chemical potential, which would correspond to half of the pion mass of finite-density QCD. 
This triggers a fictitious phase transition on each Lefschetz thimble, and the interference of complex phases among them plays an important role for the correct description of the system. 
We also show that the complex Langevin method does not work in this situation. }

\FullConference{34th annual International Symposium on Lattice Field Theory\\
		24-30 July 2016\\
		University of Southampton, UK}

\begin{document}

\section{Introduction}\label{sec:intro}

QCD at finite density is an important subject for both theoretical and experimental nuclear physics, but the first principle approach based on lattice numerical simulation suffers from the sign problem~\cite{Muroya:2003qs}. 
Especially, the sign problem of finite-density QCD becomes very severe when quark chemical potential exceeds half of the pion mass in the conventional reweighting method~\cite{Barbour:1997bh}, and we must understand properties of the functional integral itself to overcome this problem~\cite{Cohen:2003kd}. 
This article reviews the study of the sign problem of $(0+1)$-dimensional fermion model using the path integral on Lefschetz thimbles~\cite{Tanizaki:2015rda}. Complex Langevin method is also discussed for this model~\cite{Hayata:2015lzj}. 

\section{One-site Hubbard model and Silver Blaze problem}\label{sec:HubbardModel}
In order to understand the property of the Silver Blaze problem from the viewpoint of complexification, we apply the Lefschetz-thimble method to the one-site fermion model:
\be
\hat{H}={U}\hat{n}_{\up}\hat{n}_{\down}-\mu(\hat{n}_{\up}+\hat{n}_{\down}). 
\label{Eq:Hamiltonian_SS_Hubbard}
\ee
One can exactly compute its partition funtion as $Z=\mathrm{tr}[\mathrm{e}^{-\beta \hat{H}}]=1+2\mathrm{e}^{\beta\mu}+\mathrm{e}^{\beta(2\mu-U)}$. 
To show that its path-integral expression suffers from the severe sign problem, let us write down the effective action of this model with the Hubbard--Stratonovich field $\varphi$~\cite{Tanizaki:2015rda}: 
\be
S={\varphi^2\over 2U}+\psi^*\left[\p_{\tau}-\left({U\over 2}+\im \varphi +\mu\right)\right]\psi,
\ee
where $\psi=(\psi_{\uparrow},\psi_{\downarrow})$. 
The Hubbard--Stratonovich field $\varphi$ is related to the number density $n$ by $n=\mathrm{Im}\langle \varphi\rangle/U$.  The Yukawa coupling, $\im \psi^* \varphi \psi$, is complex and this is the origin of the sign problem of this model.  
This is equivalent to the $(0+1)$-dimensional massless Thirring model, and see Refs.~\cite{Pawlowski:2013pje} for relevant studies of $(0+1)$-dimensional massive Thirring model. 
The fermions do not couple with the non-zero Matsubara modes of $\varphi$, i.e., the fermion determinant is
\[
\mathrm{Det}\left[\p_{\tau}-\left({U\over 2}+\im \varphi(\tau) +\mu\right)\right]=\left(1+\mathrm{e}^{-\int_0^\beta\diff \tau\left(-U/2-\im \varphi(\tau)-\mu\right)}\right)^2. 
\]
The path integral of our interest is now reduced to an integral of zero Matsubara mode $\varphi_{\mathrm{bg}}=\int_0^\beta\diff\tau\varphi(\tau)/\beta$, and the partition function becomes
\be
Z=\sqrt{\beta\over 2\pi U}\int_{\mathbb{R}} {\diff \varphi_{\mathrm{bg}}}  \left(1+\mathrm{e}^{\beta\left(\im \varphi_{\mathrm{bg}}+\mu+U/2\right)}\right)^2\mathrm{e}^{-{\beta\varphi_{\mathrm{bg}}^2/ 2U}}. 
\label{Eq:ZeroModeIntegral_Hubbard}
\ee

Let us mention the property of the sign problem of (\ref{Eq:ZeroModeIntegral_Hubbard}) in the zero-temperature limit, and compare it with that of finite-density QCD. 
When $\mu<-U/2$, the fermion determinant converges to $1$ in the limit $\beta\to\infty$ since $\mathrm{e}^{\beta\left(\im \varphi_{\mathrm{bg}}+\mu+U/2\right)}\to 0$. This means that the sign problem at $\mu<-U/2$ is exponentially tiny at low temperatures. 
At $\mu=-U/2$, the fermion spectrum becomes gapless. The sign problem becomes severe for $\mu>-U/2$, and the mean-field number density becomes nonzero. The same thing can be discussed for the case of QCD by paying attention to the spectrum of $\gamma^0(\slashed{D}(A)+m)$~\cite{Cohen:2003kd}. The sign problem becomes severe after half of the pion mass because of fictitious pion condensation. 
Since nothing should happen at half of the pion mass because there is no such a light baryon, this  is called the baryon Silver Blaze problem. 
In our case, the lowest energy of (\ref{Eq:Hamiltonian_SS_Hubbard}) is zero, so the mean-field onset of the number density at $\mu=-U/2$ is also fictitious. 

\section{Lefschetz-thimble method} %to the repulsive Hubbard model}
\label{sec:LefschetzThimble_OneSiteHubbard}

Let us consider a generic situation that the partition function is given by
\be
Z=\int_{\mathbb{R}^n}\diff^n x~\mathrm{e}^{-S_{\mathrm{eff}}(x)},
\label{Eq:General_Expression_Partition_Function}
\ee
where $S_{\mathrm{eff}}(x)$ is a complex-valued effective action. 
The idea of the Lefschetz-thimble method is that the original integration region $\mathbb{R}^n$ is not necessarily the best option after complexifying the integration variables $x^j\mapsto z^j=x^j+\im y^j$. 
In the case of one-variable integrals, the best option would be steepest descent paths and one can deform $\mathbb{R}$ to the sum of them by using Cauchy's theorem. 
Lefschetz thimbles are higher dimensional generalization of steepest descent paths, and this notion starts to get attention in the context of the sign problem quite recently~\cite{Cristoforetti:2012su}. 

Each Lefschetz thimble is an $n$-dimensional space spanned around a saddle point $z_{\sigma}$ in $\mathbb{C}^n$ ($\sigma\in\Sigma$). 
Using the gradient flow by introducing a flow time $t$~\cite{pham1983vanishing}:
\begin{equation}
\frac{\diff {z^i}(t)}{\diff t} =   \overline{\left(\frac{\partial S_{\mathrm{eff}}(z(t))}{\partial z^i}\right)},
\label{Eq:Downward_Flow}
\end{equation}
one defines the Lefschetz thimble and its dual as 
\begin{equation}
\mathcal{J}_{\sigma}:=\left\{z(0)\,\Bigl{|}\, \lim_{t\to-\infty}z(t)=z_{\sigma}\right\}, \qquad
 \mathcal{K}_{\sigma}:=\left\{z(0)\,\Bigl{|}\, \lim_{t\to+\infty}z(t)=z_{\sigma}\right\}, 
\end{equation}
respectively. The path integral (\ref{Eq:General_Expression_Partition_Function}) is now written as 
\be
Z=\sum_{\sigma\in\Sigma}\langle \mathbb{R}^n,\mathcal{K}_{\sigma}\rangle\int_{\mathcal{J}_{\sigma}}\diff^n z~\mathrm{e}^{-S(z)}, 
\label{Eq:Lefschetz_Thimble_Decomposition}
\ee
where the coefficient $\langle \mathbb{R}^n,\mathcal{K}_{\sigma}\rangle$ is the intersection number between $\mathbb{R}^n$ and $\mathcal{K}_{\sigma}$.
The validity of this method with fermion determinant is discussed in Refs.~\cite{Tanizaki:2014tua}. 

\section{Semiclassical analysis of one-site Hubbard model using Lefschetz thimbles}\label{sec:SemiClassical_OneSiteHubbard}
Let us apply the Lefschetz-thimble method to the path integral~\eqref{Eq:ZeroModeIntegral_Hubbard}. The effective action is  
\be
S_{\mathrm{eff}}(z)={\beta\over 2U}z^2-2\ln \left(1+\exp\beta\left(\im z  +\mu+{U\over 2}\right)\right). 
\label{Eq:EffectiveAction_Hubbard}
\ee
The effective action satisfies the CK symmetry ensuring the real-valuedness of the Lefschetz-thimble decomposition manifestly \cite{Tanizaki:2015pua}. 

In the low-temperature limit $T\ll U,|\mu|$, we can find the saddle points of $S_{\mathrm{eff}}$ as~\cite{Tanizaki:2015rda}
\be
z_{m}=\im\left(\mu+{U\over 2}\right)+T\left(2\pi m +\im\ln {{3\over2}U-\mu\over {1\over 2}U+\mu}\right)+O(T^2)
\label{Eq:ApproximateSaddlePoints}
\ee 
for $m\in\mathbb{Z}$. 
If $\mu>3U/2$ or $\mu<-U/2$, we can also find another saddle point,
\be
z_{*}=
\left\{\begin{array}{ccc}
2\im U+o(T) & {\rm for} & \mu/U>{3\over 2},\\ 
0+o(T) & {\rm for} & \mu/U<-{1\over 2}.
\end{array}\right. 
\ee

\begin{figure}[t]
\begin{minipage}{.45\textwidth}
\subfloat[$\mu/U=-1$]{
\includegraphics[scale=0.45]{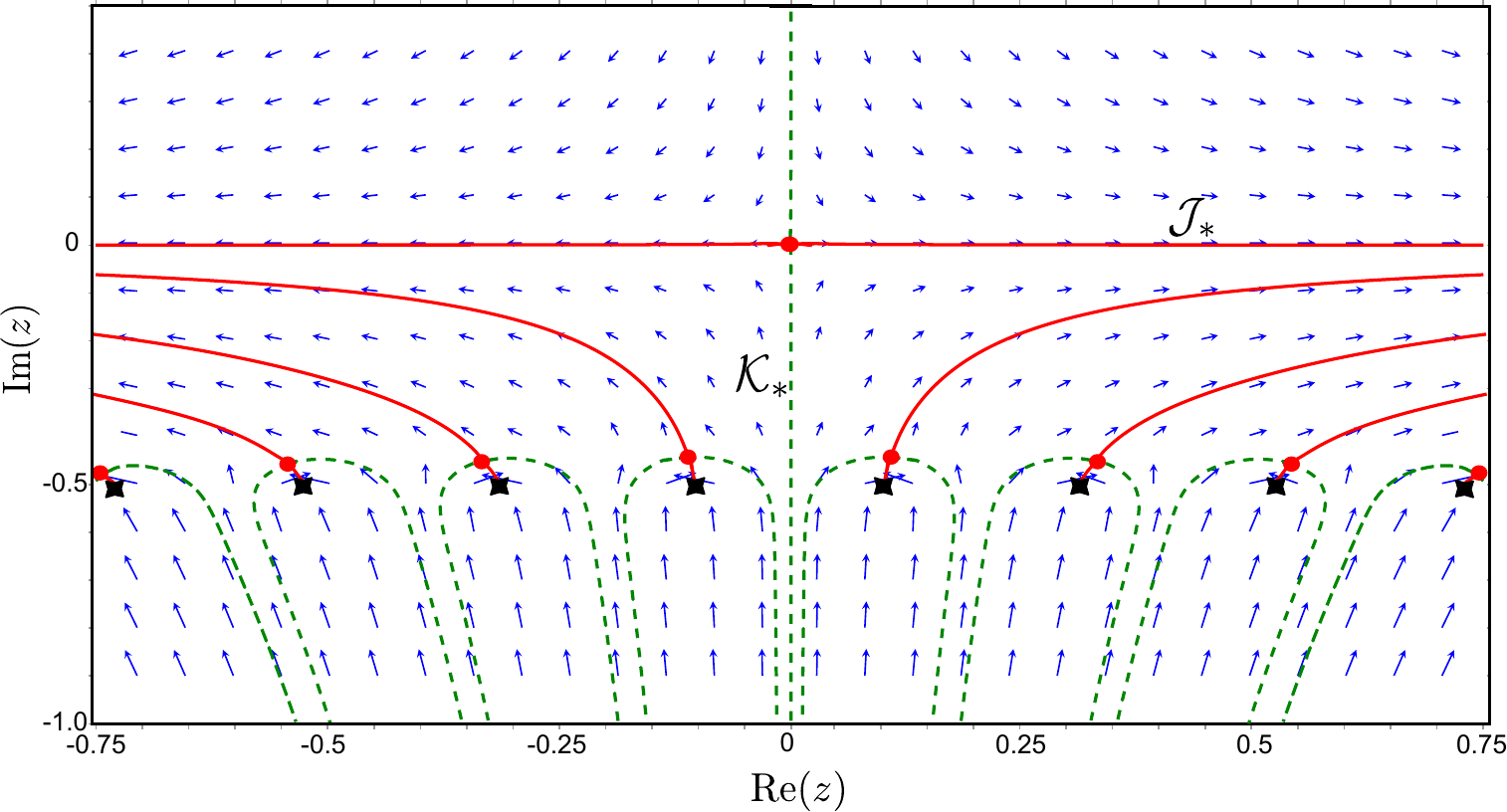}
\label{fig:Flow_EasySilverBlaze}
}
\end{minipage}\quad%\vspace{0.5em}
\begin{minipage}{.45\textwidth}
\subfloat[$\mu/U=0$]{
\includegraphics[scale=0.45]{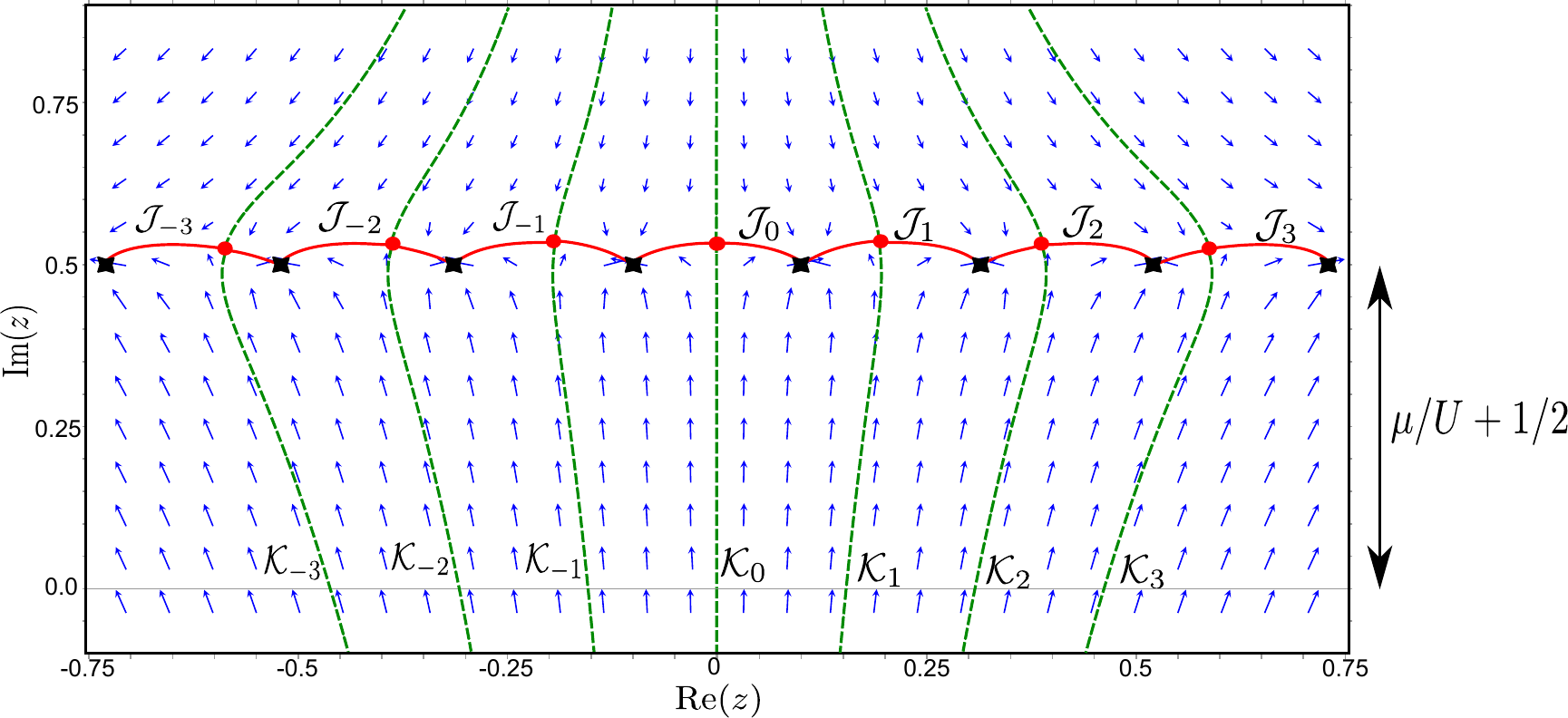}
\label{fig:Flow_DifficultSilverBlaze}
}
\end{minipage}
\caption{
Behaviors of the gradient flow equation for $\beta U=30$, $U=1$, and $\mu/U=-1$ (a) [$\mu/U=0$ (b)]~\cite{Tanizaki:2015rda}.
Star-shape black points show singular points of logarithm, and red blobs are $z_{\sigma}$. 
}
\label{fig:Flow_HubbardModel}
\end{figure}

Behaviors of the gradient flow equation \eqref{Eq:Downward_Flow} are shown in Fig.~\ref{fig:Flow_HubbardModel} with $U=1$ and $\beta U=30$. 
Fig.~\ref{fig:Flow_EasySilverBlaze} shows the flow when the chemical potential is sufficiently large and negative ($\mu/U=-1$). 
Only one Lefschetz thimble $\mathcal{J}_*$ associated with $z_*$ contributes, and $\mathcal{J}_*$ is almost identical with $\mathbb{R}$ because the sign problem is exponentially weak. 
The number density is 
\be
n\simeq {-\im z_*\over U}=0. 
\ee 

In the following, let us consider the case with the difficult Silver Blaze problem, $-1/2\lesssim \mu/U\lesssim 3/2$, at low temperatures $\beta U\gtrsim 10$. 
In this case, the behavior of the gradient flow is shown in Fig.~\ref{fig:Flow_DifficultSilverBlaze}. 
All dual thimbles $\mathcal{K}_m$ intersect with $\mathbb{R}$, which means that all the saddle points $z_{m}$ contribute to the partition function. 
If one neglects this fact and picks up the single saddle point $z_0$, the mean-field approximation is recovered to find that 
\be
n_{\mathrm{MF}}={1\over U}\mathrm{Im}(z_0)={\mu\over U}+{1\over 2}. 
\ee
This is completely wrong, since $n=0$ for $\mu<0$. 
%As $\mu$ increases from the region $\mu/U<-1/2$, the zeros of fermion determinant move and go across the real axis. This induces the Stokes phenomenon, and the structure of Lefschetz thimbles totally changes. 

We compute the classical actions $S_{\sigma}:=S_{\mathrm{eff}}(z_{\sigma})$ in the low temperature limit as 
\begin{eqnarray}
&S_0\simeq-{\beta U\over 2}\left(\frac{\mu}{U}+{1\over 2}\right)^2,
\label{Eq:ApproximateClassicalAction_OneThimble}\\
&\mathrm{Re}\,(S_m-S_0)\simeq {2\pi^2\over \beta U}m^2, 
\label{Eq:ApproximateClassicalActionRe}\\
&\mathrm{Im}\,S_m\simeq 2\pi m\left({\mu\over U}+\frac{1}{2}\right).
\label{Eq:ApproximateClassicalActionIm} 
\end{eqnarray}
According to \eqref{Eq:ApproximateClassicalActionRe}, subdominant thimbles $\mathcal{J}_m$ contribute comparably with the dominant one $\mathcal{J}_0$ for $\beta U\gg 1$ so long as $|m|(\not=0)$ is not too large.
To consider the impact of interference among multiple complex saddles, we consider the classical approximation of the partition function:
\be
Z_{\mathrm{cl}}:=\sum_{m=-\infty}^{\infty}\mathrm{e}^{-S_m}=\mathrm{e}^{-S_0(\mu)}\theta_3\left(\pi\left({\mu\over U}+{1\over 2}\right),\mathrm{e}^{-2\pi^2/\beta U}\right). 
\label{Eq:ApproximateSemiclassicalSummation}
\ee
Using this result, in the limit $\beta U\to \infty$, the number density is given as 
\be
n_{\mathrm{cl}}:={1\over \beta}{\p\over \p \mu}\ln Z_{\mathrm{cl}}
\to 
\left\{\begin{array}{cl}
2& (1<\mu/U<3/2),\\
1& (0<\mu/U<1),\\
0& (-1/2<\mu/U<0). 
\end{array}\right.
\label{eq:Number_Density_Semiclassical_T0}
\ee
This success indicates the usefulness of the semiclassical approximation even when the sign problem is severe. 

How many Lefschetz thimbles are relevant in the sum (\ref{Eq:ApproximateSemiclassicalSummation}) at a given lower temperature $\beta$? 
We propose the criterion to neglect the Lefschetz thimbles $\mathcal{J}_m$ for large $|m|$, 
$
\left|Z_{\mathrm{cl}}^{(2m+1)}-Z_{\mathrm{cl}}^{(2m-1)}\right|\ll |Z_{\mathrm{cl}}^{(2m+1)}|,  
$
where $Z_{\mathrm{cr}}^{(2m+1)}=\sum_{n=-m}^{m}\mathrm{e}^{-S_n}$. 
Solving this condition at $\mu=0$, we find that 
\be
|m|\gtrsim {\beta U\over 4\pi}
\label{Eq:NecessaryNumbersOfThimbles_Mu0}
\ee
in the limit of $\beta U\gg 1$. 
It means that we need at least $(2\lceil\beta U/4\pi\rceil +1)$ thimbles in order to describe the rapid crossover of the number density in the case of the one-site Hubbard model~\cite{Tanizaki:2015rda}. 

\begin{figure}[t]\centering
\includegraphics[scale=0.7]{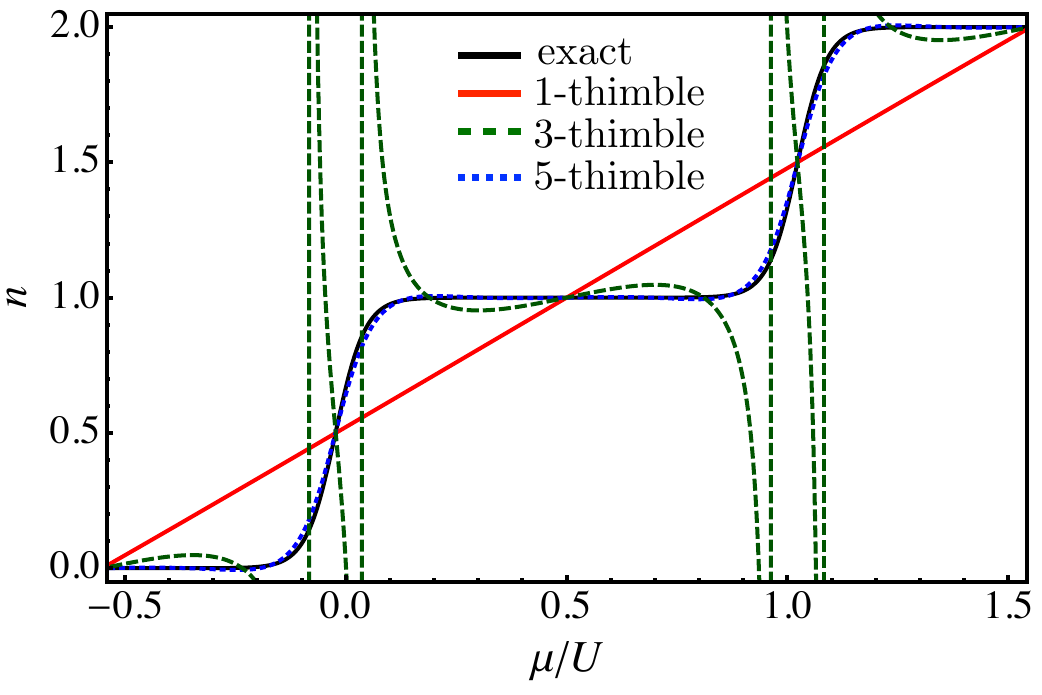}
\caption{
Number density as a function of $\mu$ at $\beta U=30$~\cite{Tanizaki:2015rda}. 1-, 3-, 5-thimble results are computed by taking into account $\mathcal{J}_0$, $\mathcal{J}_0\cup\mathcal{J}_{\pm1}$, $\mathcal{J}_0\cup\mathcal{J}_{\pm1}\cup\mathcal{J}_{\pm2}$, respectively. 
}
\label{fig:NumberDensity}
\end{figure}

By performing the exact computation of the path integral on Lefschetz thimbles, we can indeed check that above semiclassical analysis gives the correct result in this model. The result for the number density at $\beta U=30$ is shown in Fig.~\ref{fig:NumberDensity}, and we can see the impact of the interference among complex saddles. 

\section{Comment on the failure of complex Langevin method}
We can show and confirm that the complex Langevin method does not work in the one-site Hubbard model at $-1/2<\mu/U<3/2$. More generally, if there are several complex saddle points that dominantly contribute with different complex phases, we can prove that the complex Langevin method does not work when the semiclassical analysis is valid~\cite{Hayata:2015lzj}. 
Let us denote the set of dominant complex saddles by $\{z_{\delta}\}_{\delta\in\Delta}$ ($\Delta\subset\Sigma$), then the complex-Langevin expectation value $\langle\cdot\rangle_{\mathrm{CL}}$ of a holomorphic operator $O(z)$ is well approximated by 
\be
\langle O(z)\rangle_{\mathrm{CL}}=\sum_{\delta\in\Delta}c_{\delta} O(z_{\delta}). 
\ee
Here $c_{\delta}\ge 0$ because the expectation value $\langle\cdot\rangle_{\mathrm{CL}}$ is defined by the ensemble average of the stochastic process. As the hypothesis for contradiction, we suppose that the complex Langevin method gives the same result with the original path integral. Under this hypothesis, 
\bea
\langle O(z)\rangle_{\mathrm{CL}}&=&{1\over Z}\int \diff x\, \mathrm{e}^{-S_{\mathrm{eff}}(x)} O(z)
=\sum_{\sigma\in\Sigma}{\langle\mathbb{R},\mathcal{K}_{\sigma}\rangle\over Z}\int_{\mathcal{J}_{\sigma}}\diff z\, \mathrm{e}^{-S_{\mathrm{eff}}(z)} O(z)\nonumber\\
&\simeq& \sum_{\delta\in \Delta}{\langle\mathbb{R},\mathcal{K}_{\delta}\rangle\over Z}\sqrt{2\pi\over S''_{\mathrm{eff}}(z_\delta)}\mathrm{e}^{-S_{\mathrm{eff}}(z_{\delta})}O(z_{\delta}). 
\eea
This claims that $c_{\delta}={\langle\mathbb{R},\mathcal{K}_{\delta}\rangle\over Z}\sqrt{2\pi\over S''_{\mathrm{eff}}(z_\delta)}\mathrm{e}^{-S_{\mathrm{eff}}(z_{\delta})}\ge 0$ for all $\delta\in\Delta$. 
This is the contradiction, and the complex Langevin method is wrong in this situation. 

\begin{figure}[t]
\begin{minipage}{.45\textwidth}
\subfloat[Complex Langevin distribution and Lefschetz thimbles]{
\includegraphics[scale=0.45]{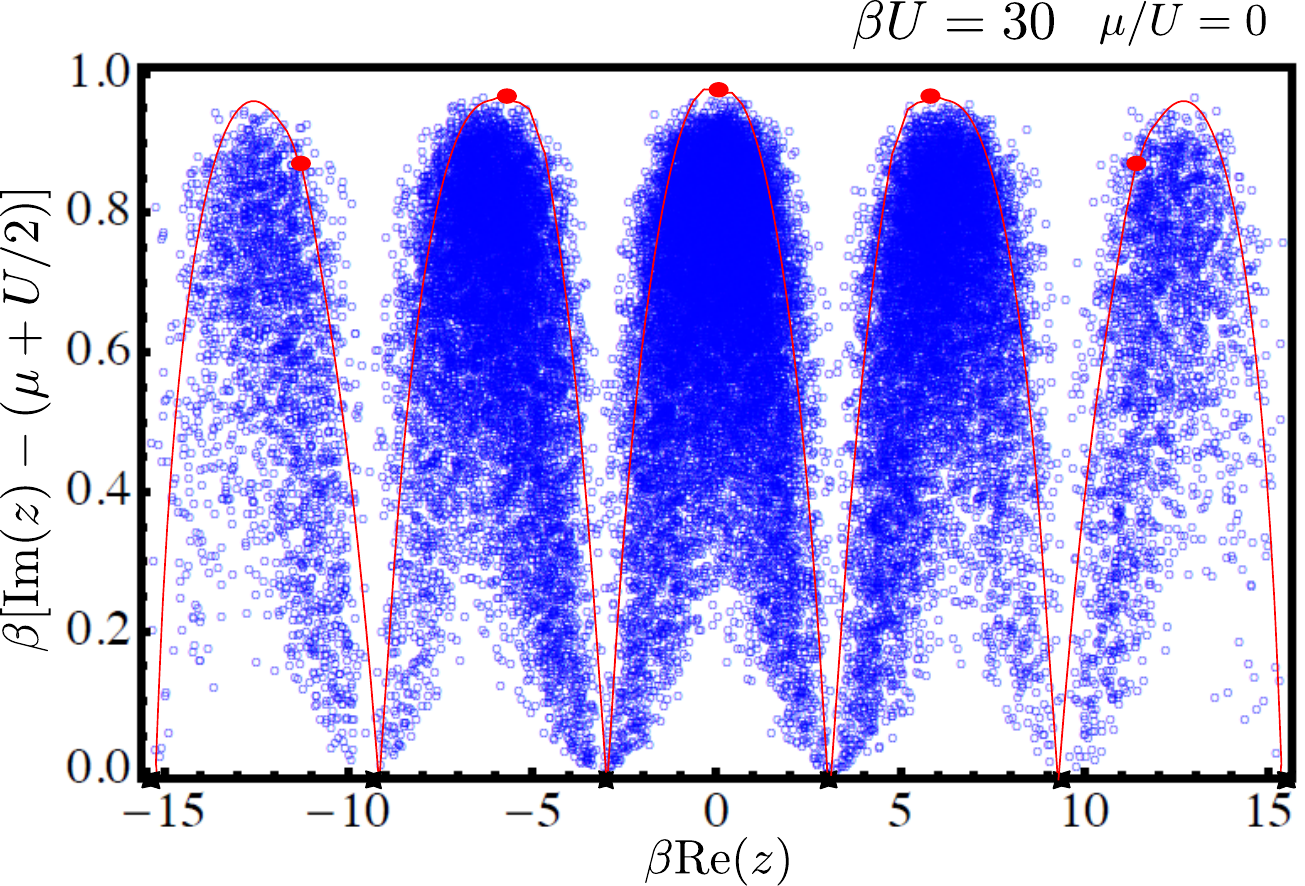}
\label{fig:CLdist}
}
\end{minipage}\quad%\vspace{0.5em}
\begin{minipage}{.45\textwidth}
\subfloat[Number density at $\beta U=30$]{
\includegraphics[scale=0.6]{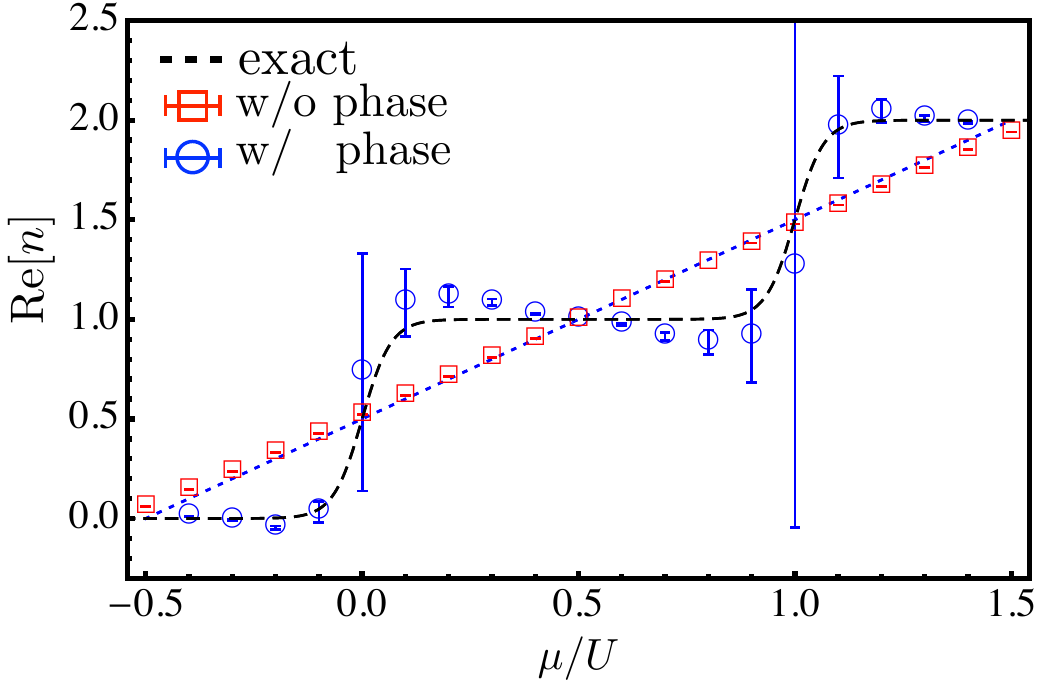}
\label{fig:CLHubbard}
}
\end{minipage}
\caption{
Results of the complex Langevin method for the one-site Hubbard model~\cite{Hayata:2015lzj}. 
}
\label{fig:CL}
\end{figure}

Let us show, in Fig.~\ref{fig:CL}, the numerical results of the complex Langevin method for the one-site Hubbard model at $\beta U=30$. Fig.~\ref{fig:CLdist} compares the complex Langevin distribution (blue circles) and profiles of Lefschetz thimbles (red curves) at $\mu=0$, and complex Langevin distribution is divided into several regions. 
Red squares of Fig.~\ref{fig:CLHubbard} show the number density by complex Langevin method, which give wrong mean-field results, $n_{\mathrm{MF}}$. 
In order to emphasize again the importance of phases, we perform the reweighting in the complex Langevin method by assigning the phase factor of Lefschetz thimbles to each local distribution of the complex Langevin method~\cite{Hayata:2015lzj}. 
The result is shown by blue circles in Fig.~\ref{fig:CLHubbard}, and the result is clearly improved by this procedure. 

\section{Summary}\label{sec:summary}

We study the one-site Hubbard model using the Lefschetz-thimble method especially at low temperatures. In the path-integral expression, this model has a sign problem and the property of the Silver Blaze problem is very similar to that of finite-density QCD with light quarks. 
At the chemical potential where the complex mean-field approximation becomes bad, Stokes phenomenon happens and the original integration cycle starts to be decomposed into several Lefschetz thimbles. 
In order to get the correct answer, the summation over Lefschetz thimbles has a significant effect, which has a milder sign problem compared with the original integration. 
It is natural to guess that the same thing happens also for finite-density QCD beyond half of the pion mass. 

We also discuss the complex Langevin method from the viewpoint of the saddle-point approximation, and we show that it is wrong in the above situation. The difference of phases among different dominant saddles bring the important physical information, but it is lost in the complex Langevin method at least in the semiclassical regime. 

\section*{Acknowledgments}
{Y.~T. is supported by the Special Postdoctoral Researchers Program of RIKEN. 
Y.~H. is partially supported by JSPS
KAKENHI Grant No. 15H03652 and 16K17716 and by RIKEN iTHES project. 
T. H. is supported by Grants-in-Aid for the fellowship of
JSPS (No:
JP16J02240)}

%\bibliographystyle{utphys}
%\bibliography{lefschetz,ref_hubbard}
\providecommand{\href}[2]{#2}\begingroup\raggedright\endgroup

\end{document}